\documentclass[12pt,a4paper]{amsart}
\usepackage{amsmath,amssymb,amsthm,verbatim,xypic}
\newtheorem{theorem}{Theorem}[section]

\theoremstyle{remark}

\theoremstyle{definition}
\newtheorem{definition}[theorem]{Definition}
\renewcommand{\phi}{\varphi}

\newcommand{\N}[1]{\left\Vert #1 \right\Vert}

\newcommand{\st}{\, : \,}

\newcommand{\mb}{\mathbb}

\newcommand{\cmax}{\otimes_{max}}

\begin{document}
\title[Joint system quantum descriptions]{Joint system quantum descriptions arising from local quantumness}

\author[Cooney]{Tom Cooney}
\email{tomcooney1@gmail.com}
\address{Departamento de An{\'a}lisis Matem{\'a}tico \\
Facultad de Matem{\'a}ticas\\
\noindent Universidad Complutense de Madrid \\
Madrid 28040\\
Spain}

\author[Junge]{Marius Junge}
\email{junge@math.uiuc.edu}
\address{Department of Mathematics\\ University of Illinois at Urbana-Champaign\\
Illinois 61801-2975\\ USA}

\author[Navascu\'es]{Miguel Navascu\'es}
\email{m.navascues@mat.ucm.es}
\address{School of Physics\\ University of Bristol \\
Bristol BS8 1TL\\U.K.}

\author[P\'erez-Garc\'{\i}a]{David P\'erez-Garc\'{\i}a}
\email{dperez@mat.ucm.es}

\address{Departamento de An{\'a}lisis Matem{\'a}tico \\
Facultad de Matem{\'a}ticas\\
\noindent Universidad Complutense de Madrid \\
Madrid 28040\\
Spain}

\author[Villanueva]{Ignacio Villanueva}
\email{ignaciov@mat.ucm.es}
\address{Departamento de An{\'a}lisis Matem{\'a}tico \\
Facultad de Matem{\'a}ticas\\
\noindent Universidad Complutense de Madrid \\
Madrid 28040\\
Spain}
\begin{abstract}
Bipartite correlations generated by non-signalling physical systems that admit a finite-dimensional local quantum description cannot exceed the quantum limits, i.e., they can always be interpreted as distant measurements of a bipartite quantum state. Here we consider the effect of dropping the assumption of finite dimensionality. Remarkably, we find that the same result holds provided that we relax the tensor structure of space-like separated measurements to mere commutativity. We argue why an extension of this result to tensor representations seems unlikely.
\end{abstract}

\maketitle

\section{Introduction}

One of the most remarkable and counterintuitive features of Quantum Mechanics is the ``spooky action at a distance" which manifests itself best through the existence of what we call non-local correlations.  These are joint probability distributions between two (or more) space-like separated parties which can not be explained by any local realistic model. 

At the same time, our physical models assume the finite speed of the propagation of light from which we infer the finite speed of the propagation of information. From this restriction we are led to believe that all physically realizable joint probability distributions between two space-like separated parties must be {\em non-signalling}, a notion we formalize below. 

We know that the three sets of local, quantum, and non-signalling probability distributions are each strictly contained in the next. Much work has been done in understanding the consequences of this fact, both from a practical and a foundational point of view.  

In particular, we seek explanations for the fact that the amount of non-locality of quantum distributions is not enough to saturate the set of non-signalling distributions. Several physical principles have been proposed to explain this discrepancy between quantum and non-signalling predictions, like Non-Trivial Communication Complexity \cite{NTCC}, Information Causality \cite{IC} and Macroscopic Locality \cite{ML}.

Recently, in the paper \cite{BBBEW} (see also \cite{acin}), the authors address these types of questions from the following point of view: to what extent does ``local quantumness" plus the non-signalling condition restricts the ``amount of non-locality" of joint distributions?

They showed that {\em for finite dimensional local systems} assuming local quantumness plus non-signalling implies that every  physical preparation of the joint system can be simulated by a quantum description, in the sense that all possible bipartite probability distributions can be explained with this simulation (see Theorem \ref{teo0} below). 

The assumption of finite dimensionality is crucial in their proof. From a foundational point of view, it is important to determine whether a similar result holds for infinite dimensional systems, since so far infinite dimensions are needed to model many important physical situations. Indeed, in non-relativistic quantum mechanics, the positions and momenta of a finite set of particles must satisfy a number of polynomial identities -the canonical commutation relations- that do not admit a finite dimensional representation. Hence the possible states of such particles are represented as rays of a separable infinite-dimensional Hilbert space \cite{quantum}. Likewise, dynamical equations and operators in Quantum Field Theory are defined in infinite dimensional (generally non-separable) Hilbert spaces \cite{QFT}. It is also worth noticing that, although most protocols and algorithms in Quantum Information Theory were initially conceived for finite dimensional systems, their actual experimental realizations typically involve infinite dimensions \cite{CV_crypt,CV_tel}. For instance, in most quantum computation experiments, two internal states of an ion are used to model a qubit \cite{ion_trap}. This approximation breaks at temperatures of the order of 1 K, though, as thermal fluctuations drive the system along an infinite-dimensional state space.

In this paper we show that a result analogous to \cite{BBBEW}, \cite{acin} does hold true in infinite dimensions, see Theorem \ref{teo2} below. However, in our proof we need to use the ``commutative" (as opposed to ``tensor") description of joint quantum systems to provide the desired simulation. Indeed, we give reasons below why we find it quite unlikely that such a simulation in the tensor description always exists. 

The result requires the use of involved techniques developed in von Neumann algebra theory or, alternatively, in non-commutative $L_p$-spaces theory. 

Finally, we provide an additional contribution to this problem: as it turns out, both in the finite and infinite dimensional case, physical preparations can be simulated via a pair of local transformations $\nu_A,\nu_B$ which map \emph{individual} measurement operators to individual measurement operators (as opposed to more general local transformations that would map \emph{complete} sets of measurement operators to complete sets of measurement operators). In section \ref{projec} we show that, even in the finite-dimensional case, such models cannot provide a quantum representation of general physical preparations if we further demand that $\nu_A,\nu_B$ map projectors to projectors.

We formalize now these ideas. We will consider two space-like separated physical systems $A$ and $B$ (Alice and Bob's systems). We assume that, separately, each of them can be described by Quantum Mechanics. We call this {\em Local Quantumness}:

(LQ) {\em Local Quantumness}: When considered separately, Alice's system is associated to a unital C$^*$-algebra $A$ 
such that, for every physical preparation of the system, there exists a state $\rho
\in \mathcal S(A)$ with the following property: for every POVM $\{Q_a\}_{a=1}^N\subset A$, there exists a physical measurement such that $p(a)=\rho(Q_a)$. Here $\mathcal S(A)$ denotes the states on $A$, the positive linear functionals on $A$ of norm 1. Similarly, Bob's system is associated to a unital C$^*$-algebra $B$.

We want to explore to what extent Local Quantumness forces the joint system to admit also a quantum mechanical description. Therefore, we do not assume that the joint system can be described by Quantum Mechanics. But, as explained above, we do assume that the joint system is  {\em non-signalling}. 

(NS) {\em Non-signalling}: When considered jointly, the
non-signalling principle holds. That is, for every physical
preparation of the joint system, and for every set of questions
$x,y=1,\ldots, M$ with possible answers $a,b=1,\ldots, N$, the joint
probability distribution $p(a,b|x,y)$ satisfies
\begin{align*}
\sum_a P(a,b|x,y)=p(b|y) \text{ is independent of } x,\\
\sum_b P(a,b|x,y)=p(a|x) \text{ is independent of } y.
\end{align*}

The physical preparations of the system verifying (LQ) and (NS) will
be called {\em valid preparations}. If all the preparations of the
system are valid preparations, we say that  our system is {\em
locally quantum} and {\em non-signalling}.

In a recent paper \cite{BBBEW}, Barnum, Beigi, Boixo, Elliott and
Wehner proved that if $A$ and $B$ are the bounded operators on
finite dimensional Hilbert spaces $H_A$, $H_B$, then every  valid
preparation on the joint system can be simulated by a state in  the
Hilbert space $H_A\otimes H_B$.

Let us formalize that statement.  We denote the
positive self-adjoint elements of $B(H_A)$ bounded from above by the
identity $1$ by $B(H_A)_{sa}^{+,1}$. Any valid preparation can be described by a function $\omega:B(H_A)_{sa}^{+,1}\times
B(H_B)_{sa}^{+,1}\to[0,1]$ with the restrictions
following from properties (LQ) and (NS) above.

With this language, the main result of \cite{BBBEW} can be stated as follows.

\begin{theorem}\label{teo0}
For every valid preparation $\omega$ as above, there exists a state
$\rho$ in $B(H_A\otimes H_B)$ and mappings $\nu_A:
B(H_A)_{sa}^{+,1}\to B(H_A)_{sa}^{+,1}$, $\nu_B:
B(H_B)_{sa}^{+,1}\to B(H_B)_{sa}^{+,1}$ carrying
measurement systems into measurement systems such that for every
$(Q_a, R_b) \in B(H_A)_{sa}^{+,1}\times B(H_B)_{sa}^{+,1}$,
$$\omega(Q_a, R_b)= tr(\rho \nu_A(Q_a)\otimes \nu_B(R_b)).$$
\end{theorem}

\smallskip

In this paper we study whether this result remains true if we
remove the hypothesis on the finite dimensionality of the two
subsystems. We prove that an analogous result is true if we allow the joint system to be simulated in the so-called {\em commutative paradigm}. That is, for every possible choice of the local  C$^*$-algebras $A$ and
$B$,  every valid preparation of the joint system can be simulated
by a state on a Hilbert space $H$ such that $B(H)$ contains $A$ and
$B$ as subalgebras commuting with each other.

The precise statement now is

\begin{theorem}\label{teo2}
For every choice of the local C$^*$-algebras $A$ and $B$ and for
every valid preparation $\omega$ on the joint system, there exists a
Hilbert space $H$, a state $\rho$ on $B(H)$, and mappings  $\nu_A:
A_{sa}^{+,1}\to B(H)_{sa}^{+,1}$, $\nu_B: B_{sa}^{+,1}\to B(H)_{sa}^{+,1}$ carrying
measurement systems into measurement systems such that $\nu_A(A)$
commutes with $\nu_B(B)$ and such that for every $(Q_a, R_b) \in
A_{sa}^{+,1}\times B_{sa}^{+,1}$,
$$\omega(Q_a, R_b)= \rho (\nu_A(Q_a) \nu_B(R_b)).$$
\end{theorem}

At first sight it might seem as if the natural extension of the finite dimensional version should provide a simulation of the joint system in the {\em tensor paradigm}. That is, in Theorem \ref{teo2} above we would want Hilbert spaces $H_A$ and $H_B$, a state $\rho_{AB}$ on $B(H_A\otimes H_B)$ and  mappings $\nu_A:
A_{sa}^{+,1}\to B(H_A)_{sa}^{+,1}$, $\nu_B: B_{sa}^{+,1}\to B(H_B)_{sa}^{+,1}$.

Actually, the existence of such simulation would imply a resolution of a very strong version of Tsirelson's problem, which leads us to believe that such a result cannot be true. We briefly recall Tsirelson's problem and explain this.  

We suppose a physical system composed of two space-like separated subsystems A and B. On this system we consider a physical experiment which  can be repeated an arbitrary number of times. This experiment has a fixed finite  number of possible inputs and outputs ($N$ and $M$) to each party. Upon many realizations of the experiment, for any pair of inputs $(x,y)$ and any pair of outputs $(a, b)$ we obtain a probability distribution $P=P(a,b|x,y)$. Following \cite{Tsirelson} we will \textbf{call} such a probability distribution a {\em behaviour}.  

We say that a behaviour belongs to the {\em tensor paradigm} if there exist two Hilbert spaces $H_A$, $H_B$, a state $\rho_{AB}\in S(B(H_A\otimes H_B))$ and , for all inputs $x$ and $y$, two sets of measurement operators  $\{E_a^x\in B(H_A)\}$, $\{F_b^y\in B(H_B)\}$ such that $P(a,b|x,y)=tr(\rho_{AB} E_a^x\otimes F_b^y)$.

We say that a behaviour belongs to the {\em commutative paradigm} if there exists a Hilbert space  $H$, a state $\rho\in S_1(H)$ and, for all inputs  $x$ and $y$, two sets of measurement operators  $\{E_a^x, F_b^y \in B(H)\}$,  such that $P(a,b|x,y)=tr(\rho E_a^xF_b^y)$ and $[ E_a^x, F_b^y ]=0$ for all $a,b,x,y$.   

Clearly, every behaviour in the tensor paradigm also belongs to the commutative paradigm. The question of whether every behaviour in the commutative paradigm can be arbitrarily well approximated by a behaviour in the tensor paradigm is usually called Tsirelson's problem. It is well known that both paradigms coincide if the Hilbert spaces involved have finite dimension. 

We note that if a version of Theorem \ref{teo2} providing  a simulation in the tensor paradigm would be true, then we would easily prove that, for every state in the commutative description, there would exist a corresponding state in the tensor description giving rise to the same behaviours. This would be a ``uniform" (in the number of inputs and outputs and in the choice of measurement operators)  version of Tsirelson's problem which seems highly unlikely to be true.

\section{Mathematical tools}\label{Math}
Given two C$^*$-algebras $A$, $B$ we consider the tensor product
$A\otimes B$. This tensor product has a natural algebra structure
with the product given by $(a\otimes b)(c\otimes d)=ac\otimes bd$.
If we want the product $A\otimes B$ to be a C$^*$ algebra, we need
to endow it with a norm. In general, there is more than one norm $\alpha$ on
$A\otimes B$ that turns its completion $A\otimes_\alpha
B$ into a C$^*$-algebra. The biggest and smallest C$^*$-algebra
norms on $A\otimes B$ are called $max$ and $min$ respectively. They
are defined in the following way:

\begin{definition}
Let $x  = \sum_{i=1}^n a_i \otimes b_i \in A \otimes B$. Then its maximal C$^*$-tensor norm is 
\[
\N{x}_{max} = \sup\left\{\N{\sum_{i=1}^n \pi_1(a_i) \pi_2(b_i)}_{B(H)} \right\},
\]
where the supremum is taken over all possible Hilbert spaces $H$ and $*$-homomorphisms $\pi_1:A \to B(H)$, $\pi_2:B\to B(H)$ such that $\pi_1(A)$ and $\pi_2(B)$ commute. The \emph{maximal tensor product} $A \otimes_{max} B$ is the completion of $A \otimes B$ with respect to the norm $\N{\cdot}_{max}$.
\end{definition}

\begin{definition}
Let $A$ and $B$ be embedded in $B(H_1)$ and $B(H_2)$ respectively. Given $x  = \sum_{i=1}^n a_i \otimes b_i \in A \otimes B$, its minimal tensor product norm is $\N{x}_{min}=\N{x}_{B(H_1 \otimes H_2)}$. (This norm can be shown to be independent of the choice of embeddings of $A$ and $B$ into $B(H_1)$ and $B(H_2)$.) The \emph{minimal tensor product} $A \otimes_{min} B$ is the completion of $A \otimes B$ with respect to the norm $\N{\cdot}_{min}$. 
\end{definition}

Whether the minimal and maximal tensor norms coincide for a given pair $A,B$ of C$^*$-algebras is always
a relevant question. It is known that for {\em nuclear}
C$^*$-algebras (in particular $M_N$) $min$ and $max$ coincide
\cite{pisier}. It is also known that, for an infinite dimensional
Hilbert space $H$, $B(H)\otimes_{min}B(H)\not =
B(H)\otimes_{max}B(H)$ (\cite{pisier}). The question of whether $min$
and $max$ coincide for $C(F_\infty)\otimes C(F_\infty)$ is
(equivalent to) Connes' conjecture, and it is again related to Tsirelson's
problem \cite{ConnesTsirelson1, ConnesTsirelson2}.

\subsection{Relation between complete positivity and the max norm}\label{cpsubsection}
Let $A$ and $B$ be C$^*$-algebras. We say that a map $T: A \to B$ is \emph{completely positive} if the maps 
$id_n \otimes T: M_n(A) \to M_n(B)$ are positive for all $n \in \mathbb N$. 
We can also define a matrix order on the dual of a C$^*$-algebra; a matrix 
$(\phi_{ij}) \in M_n(B^*)$ is positive if for all positive elements $(y_{ij})\in M_n(B)$,
\[
\sum_{i,j} \langle \phi_{ij}, y_{ij} \rangle \geq 0.
\]
This allows us to define complete positivity for maps from a C$^*$-algebra to the dual of a C$^*$-algebra. We consider an operator
$\hat \omega:A\to B^*$, where $A$ and $B$ are  C$^*$ algebras. We say that $\hat \omega$ is {\em positive} if, for
all positive elements $a\in A$, $b\in B$, we have that  $\hat\omega(a)(b)\geq 0$. We say
that $\hat \omega$ is completely positive if $id_n \otimes \hat \omega$ is positive for all $n$. More explicitly, $\hat \omega$ is completely positive if, for every $n\in \mathbb N$ and
for all positive elements $x\in M_n(A)$, $y\in M_n(B)$,

$$\sum_{i,j} \langle \hat \omega(x_{ij}) , y_{ij} \rangle \geq 0.$$

We will use the following result \cite[Theorem 11.2]{pisier}:

\begin{theorem}\label{PisierTheorem}
Let $\omega :A \otimes B \to  \mathbb C$ be a linear
form and let $\hat \omega: A \to B^*$ be the corresponding linear
map. The following are equivalent:
\begin{enumerate}
\item $\omega$ extends to a positive linear form in the unit ball of $(A \otimes_{max} B)^*$.
\item $\hat \omega$ is a completely positive map and $\|\hat \omega\|\leq 1$.
\end{enumerate}
\end{theorem}

\subsection{Inclusion of a Von Neumann algebra $M$ into its predual $M_*$}\label{inclusion}
A von Neumann algebra $M \subseteq B(H)$ is a C$^*$-algebra that is closed in the weak$^*$-topology on $B(H)$. A linear functional $\phi:M \to \mb C$ is said to be \emph{normal} if it is weak$^*$-continuous. The set of normal linear functionals on $M$ is denoted by $M_*$ and satisfies $M=(M_*)^*$ (when $M_*$ is given the Banach space structure induced by the inclusion $M_* \subseteq M^*$); it is thus known as the \emph{predual} of $M$. In fact, von Neumann algebras can be abstractly characterized as the C$^*$-algebras which are the duals of some Banach space. A state $\phi:M \to \mb C$ is said to be faithful if $\phi(x^*x)=0$ implies that $x=0$. For further details about operator algebras, see \cite{takesaki1}.

Let $(M,\phi)$ be a von Neumann algebra together with a normal
faithful state $\phi$ and let $M_*$ denote the predual of $M$. We use the GNS construction associated with $\phi$
to represent $M$ on a Hilbert space $H$. We denote the inclusion of $M$ into $H$ by $\Lambda$. We will use the contractive, linear,
positivity-preserving inclusion of $M$ into $M_*$ used in
\cite{kosaki} and \cite{terpinterp}. Readers unfamiliar with the modular theory of von Neumann algebras may find it helpful to read Subsection \ref{FiniteDimensional} where this inclusion is discussed for the special case of a state on $M_n(\mathbb C)$.

Denote by $S$ the closure of the map $\Lambda(x) \mapsto \Lambda(x^*)$. Let
$S=J\Delta^{1/2}$ be the polar decomposition of $S$. Here $J$ is a
conjugate-linear, isometric involution and $\Delta$ is a linear,
positive, self-adjoint, non-singular operator on $H$; they are called the \emph{modular conjugation} and \emph{modular operator} respectively of $\phi$. Let $\phi_x \in
M_*$ be the linear functional satisfying
\[
\langle \phi_x, y \rangle = \left( \Lambda(x)\mid J \Lambda(y) \right ),
\]
for all $y\in M$. As $J$ is an isometry, $|\langle \phi_x, y
\rangle|\leq \N{x}\N{y}$, and the inclusion $M \hookrightarrow M_*$
is thus contractive. Here $(v|w)$ denotes the scalar product of vectors $v,w$; in physicist's notation, $(v|w)\equiv\langle w| v\rangle$.

For those familiar with spatial noncommutative $L_p$-spaces (see, for example, \cite{terpinterp}), 
this inclusion corresponds to the inclusion $x \mapsto d^{1/2}xd^{1/2}$, where $d$ is the (unbounded) 
density operator associated to the state $\phi$.

By Proposition 4 in \cite{terpinterp}, we have that
\[
\langle \phi_{x^*x}, y \rangle = \left( y J\Lambda(x)\mid J \Lambda(x)
\right),
\]
which makes it clear that the embedding $x \mapsto \phi_x$ is
positivity-preserving. It also follows that
$\N{\phi_x}_{M_*}=\phi(x)$ for $x \geq 0$. As elements in $\{JyJ\st
y\in M\}$ commute with the elements of $M$ and $J\Lambda(1)=\Lambda(1)$ (see
\cite{takesaki2} for further details), we also have that for $x,y
\in M$,
\[
\langle \phi_x, y^*y \rangle = \left( x J\Lambda(y) \mid J\Lambda(y)\right).
\]
From this it follows that if $\phi_x\geq 0$, then $x \geq 0$.

For all $x,y \in M$, we have that
\begin{align}\label{definitionv}
\langle \phi_x , y \rangle = \langle x ,\phi_y \rangle, \quad y \in
M,
\end{align}
by Proposition 6 in \cite{terpinterp}.

\section{Proof of Theorem \ref{teo2}}
In this section we prove our main result, Theorem \ref{teo2}. Since the proof is mathematically involved, for the sake of readability we first write it in detail for the special case of $A=B(H_A)$ and $B=B(H_B)$, where $H_A$ and $H_B$ are finite-dimensional Hilbert spaces. This finite-dimensional situation admits a much simpler proof, but we ``expand it" to provide intuition for the constructions which we will need later in the  general (infinite-dimensional) case.

\subsection{Proof of the finite-dimensional case}\label{FiniteDimensional}
As in \cite{BBBEW}, it follows from (LC) and (NS) that any valid preparation can be described 
by a separately finitely additive function $\omega:A^{+,1}_{sa} \times B^{+,1}_{sa} \to [0,1]$. We can then use 
the bilinear version of Gleason's Theorem to prove that the above mentioned $\omega$ extends 
to a bilinear functional $\omega:A \times B \to \mathbb C$, which is positive on pure tensors (that is, for every 
$(Q , R ) \in A_+ \times B_+$, we have $\omega(Q,R)\geq 0$).

Conversely, consider a unital bilinear functional $\omega:A\times
B\to \mathbb C$. If $\omega$ is  positive on pure
tensors and $\omega(Q_a, R_b)\leq 1$ for every $(Q_a, R_b) \in
A_{sa}^{+,1}\times B_{sa}^{+,1}$, it can be associated to a valid
preparation, since it can only give rise to locally quantum, non-signalling probability distributions.

Equivalently, each valid preparation corresponds to a map
\[
\hat \omega:A \to B^* \text{ with } \hat \omega(a)(b)=\omega(a,b)
\]
satisfying
\[
\omega(a,b) \geq 0 \text{ for } a \geq 0, b \geq 0.
\]
We denote $\hat\omega(1)$ by $\phi$. As $B=B(H_B)$, there exists a positive operator $d$ on $H_B$ such that $\phi(x)=tr(d^{1/2}xd^{1/2})$, for all $x \in B$. By replacing $H_B$ with $pH_B$ where $p$ is the support of $d$, we assume that $\phi$ is faithful or, equivalently, that $d$ is invertible. 

In order to motivate the proof in the infinite-dimensional case, we discuss the modular theory of the pair $(B(H_B),\phi)$. We begin by constructing the GNS representation of $B(H_B)$ with respect to $\phi$. Let $H_\phi$ denote the Hilbert space obtained from $B(H_B)$ by defining the following inner product on $B(H_B)$, 
\[
( \Lambda(x) \mid \Lambda(y))_\phi=\phi(y^*x)=tr(d^{1/2}y^*xd^{1/2}), \qquad x,y \in B(H_B),
\]
where $\Lambda$ denotes the inclusion of $B(H_B)$ into $H_\phi$. Clearly, we have $\Lambda(x)=xd^{1/2}$, when considered as a matrix in $B(H_B)$. We obtain a $*$-representation of $B(H_B)$ on $H_\phi$ by letting $B(H_B)$ act by left multiplication,
\[
x\Lambda(y)=\Lambda(xy), \qquad x,y \in B(H_B).
\]
Let $S:H_\phi \to H_\phi$ denote the conjugate linear map $S\Lambda(x)=\Lambda(x^*)$, for $x \in B(H_B)$. This map has a polar decomposition $S=J \Delta^{1/2}$, where $J$ is a conjugate-linear isometry on $H_\phi$ satisfying $J^2=1$, and $\Delta$ is a linear, positive, invertible operator on $H_\phi$. The operator $J$ is the \emph{modular conjugation} of $\phi$ and $\Delta$ is its \emph{modular operator}. In this situation, we have
\[
J \Lambda(x)=\Lambda(d^{1/2}x^*d^{-1/2}), \qquad \Delta \Lambda(x)=\Lambda(dxd^{-1}), \qquad x \in B(H_B).
\]
Using the modular conjugation, we can write down an inclusion $x \mapsto \phi_x$ of $B(H_B)$ into $B(H_B)^*$ determined by 
\begin{align}\label{FiniteDimensionalInclusion}
\phi_{x}(y)= (  \Lambda(x) \mid J \Lambda(y) )_\phi, \qquad x,y \in B(H_B).
\end{align}
In this situation, we have $\phi_{x}(y)=tr( d^{1/2}x d^{1/2} y)$, so $\phi_{x}$ corresponds to the operator $d^{1/2}xd^{1/2}$. We thus have
\[
\{ d^{1/2} x d^{1/2} \st x \in B(H_B) \} = S_1(H_B) \simeq B(H_B)^* = \{ \phi_x \st x \in B(H_B) \}.
\]

We now return to considering the map $\hat \omega: B(H_A) \to B(H_B)^*\simeq S_1(H_B)$. This map can be factorized through $B(H_B)$ as follows: 
\begin{align*}
u: B(H_A) \to B(H_B),  & \qquad u(x)=d^{-1/2} \hat \omega(x) d^{-1/2},\\
v: B(H_B) \to S_1(H_B), & \qquad v(y)=d^{1/2} y d^{1/2},
\end{align*}
for $x \in B(H_A)$ and $y \in B(H_B)$. As $\hat \omega(1)=d$, we have that $u$ is unital and thus $u$ maps measurement systems to measurement systems. The map $v$ is also clearly completely positive in the sense of the definition in Section \ref{Math}. It is trivial that $\hat \omega = vu$. One could now apply Theorem \ref{PisierTheorem} to obtain Theorem \ref{teo0} but we continue in order to indicate how this can be generalized to the infinite-dimensional case.

Written in terms of operators, if $\hat \omega (x) =d^{1/2}yd^{1/2}$, this factorization is 
\[
\xymatrix{
x \in B(H_B)\ar[rr]^{\hat \omega} \ar[dr]_{d^{-1/2} \hat \omega( \cdot ) d^{-1/2}} && d^{1/2}yd^{1/2}\\
 & y \ar[ur]_{d^{1/2} \cdot d^{1/2}} & \\
}.
\]
However, $v$ and $u$ can also be defined in terms of linear functionals, in terms of $B(H_B)^*$ rather than $S_1(H_B)$. As
$B(H_B)^* = \{ \phi_x \st x \in B(H_B) \}$, we have that $\hat \omega(x) = \phi_{u(x)}$, for some $u(x) \in B(H_B)$. We then have that 
\[
\langle \hat \omega(x), y \rangle = tr(y d^{1/2} u(x) d^{1/2} ) = tr(d^{1/2} y d^{1/2} u(x))=\langle \phi_y, u(x) \rangle.
\]
Using the known element $\hat \omega(x)$ and the above equation, $u(x)$ can be identified with a linear functional in $(B(H_B)^*)^*=B(H_B)$.

The map $v:B(H_B) \to B(H_B)^*$ is given by $v(x)=\phi_x$ and we have
\[
vu(x)=\phi_{u(x)}=\hat \omega(x), \qquad x \in B(H_A).
\]
It is this approach that will be generalized to the case where the local systems are described by unital C$^*$-algebras.

The map $v:B(H_B) \to B(H_B)^*$ is completely positive and it follows from (\ref{FiniteDimensionalInclusion}) that it is also contractive. Thus by Theorem \ref{PisierTheorem}, the corresponding linear functional $\hat v:B(H_B) \otimes B(H_B) \to \mb C$ is max-continuous. Thus there exists a Hilbert space $K$, a state $\rho \in B(K)^*$, and commuting representations $\pi_1:B(H_B) \to B(K)$, $\pi_2:B(H_B) \to B(K)$ such that 
\[
\rho(\pi_1(x)\pi_2(y))=\hat v(x \otimes y), \qquad x,y \in B(H_B).
\]
It thus follows that 
\[
\rho( \pi_1(u(x)) \pi_2(y))=\omega(x,y), \qquad x \in B(H_A), y \in B(H_B),
\]
where $\pi_1 \circ u$ and $\pi_2$ map measurement systems to measurement systems. As $H_B$ is finite-dimensional (and thus $B(H_B)$ is nuclear), we can take $K=H_B \otimes H_B$ and recover Theorem \ref{teo0}.

\subsection{Proof of the general case} Let $A$ and $B$ be unital C$^*$-algebras
describing the local systems of Alice and Bob. Let $\omega$ be a
valid preparation for the joint system. As before, $\omega$ can be associated with
a bilinear functional $\omega:A \times B \longrightarrow \mathbb C$ which is
positive on pure tensors and such that $\omega(Q_a, R_b)\leq 1$ for
every $(Q_a, R_b)\in A_{sa}^{+,1}\times B_{sa}^{+,1}$. It follows
from the definition of the maximal tensor norm in C$^*$-algebras that
our result would be true if $\omega$ were continuous when
considered as a linear form on $A\cmax B$. In general, this is
false; not every positive on pure tensors
functional is max continuous. The idea of our proof is to factor the
operator $\hat \omega:A\longrightarrow B^*$ associated to $\omega$ as
\begin{eqnarray*}
\hat \omega:A&   \longrightarrow & B^*\\
u\searrow & &  \!\!\!\nearrow v\\
& M &
\end{eqnarray*}
where $u$ carries measurement systems to measurement systems and $v$
is completely positive (which implies that the functional $\omega_u:
M\cmax B\longrightarrow \mathbb C$ associated to $u$ is continuous).

\smallskip

We now provide further details. First we note that the
operator $\hat \omega:A\to B^*$ associated to $\omega$ is {\em positive}. 
We now identify $B^{**}$ with the universal enveloping
von Neumann algebra of $B$ (see Section III.2 of \cite{takesaki1}
for details) and $B^*$ with the predual of this von Neumann algebra.

As $\omega(1,1)=1$, we have that $\hat \omega(1)=\phi$ is a state, a
positive linear functional of norm one in $B^*$. Let $p\in B^{**}$
denote the support projection of $\phi$ and $M$ the von Neumann
algebra $pB^{**}p$. By Lemma 4.1 and following in \cite{takesaki1},
we see that $p\phi p=\phi$ (where $(p \phi p)(x)=\phi(pxp)$). If $x
\in A$, $0 \leq x \leq 1$, then $\hat \omega(x)\leq \phi$. It follows that the
support of $\hat \omega(x) \leq p$ and that $p\hat \omega(x)p=\hat \omega(x)$ for all $x \in A$.
We can thus take  $\hat \omega(A)$ to be contained in $M_*$, the normal linear
functionals on $M$ and doing so does not change the probabilities
assigned to sets of measurement operators. As the support of $\phi$
is $p$, the state $\phi$ is faithful on $M$. We represent $M$ on
a Hilbert space $H$ using the GNS representation with respect to
$\phi$.

We show now that $\hat \omega$ can be factorized as $vu$ where $u:A \to M$ is
positive and unital and $v:M \to M_*$ is completely positive. The
definition of $u$ and $v$ is motivated by our discussion of the finite-dimensional case. It can also be motivated using the language of noncommutative $L_p$-spaces. Let $d$ denote the density operator
associated with the state $\phi$ (as in \cite{terpinterp}). The map $\phi_x \mapsto d^{1/2}xd^{1/2}$ extends to an isometric isomorphism between the predual $M_*$ and $L_1(M)$, a space consisting of certain (unbounded) operators on the Hilbert space $H$.  

We obtain a map $\tilde \omega: A \to L_1(M)$ from the map $\hat \omega:A \to
M_*$ by using the isometric isomorphism between $L_1(M)$ and
$M_*$. We now motivate the definitions of $u$ and $v$ by the
following formal expressions, where
\begin{align*}
u(x) & = d^{-1/2} \tilde \omega(x) d^{-1/2}, \quad x \in A,\\
v(y) & = d^{1/2} y d^{1/2}, \quad y \in M.
\end{align*}
As $\tilde \omega(1)=d$, it is intuitively obvious that $u$ is positive and
unital and that $v$ is completely positive. This will now be shown
rigorously (without using the language of noncommutative $L_p$-spaces).

The map $v:M \to M_*$ is the inclusion discussed in Section
\ref{inclusion}. This map is linear, contractive, and completely positive,
i.e., it satisfies condition (2) of Theorem \ref{PisierTheorem}; we
have already noted that it is contractive so it remains to show that
it is completely positive. Let $tr$ denote the canonical normalized
trace on $M_n(\mathbb C)$. We then have that $tr \otimes \phi$ is a
normal faithful state on $M_n(M)$. In the same way as before, we
have a positivity-preserving inclusion of $M_n(M)$ into $(M_n(M))_*$.  Thus if $x =
(x_{ij}) \in M_n(M)_+$, we have that $(tr \otimes \phi)_x$ is a
positive element in $(M_n(M))_*$ and thus $v$ is a completely positive map.

Let $x \in A$, $0 \leq
x \leq 1$, and $y \in M$, $y\geq 0$. We seek to identify $\hat \omega(x)$ with a linear functional  $u(x)$ in  $(M_*)^*=M$. To do so, we make the following definition:
\begin{align}\label{definitionu}
\langle u(x), \phi_y \rangle = \langle y, \hat \omega(x)\rangle \text{ for all }y\in M.
\end{align}
As $\hat \omega(x) \leq \hat \omega(1)=\phi$, we
then have that
\begin{align}\label{happy}
\langle u(x), \phi_y \rangle =\langle y,\hat \omega(x) \rangle \leq \langle y, \phi \rangle =
\N{\phi_y}_{M_*}.
\end{align}
By Corollary 5(2) in \cite{terpinterp}, we have that $\{ \phi_y : y
\in M\}$ is dense in $M_*$; this implies that $u(x)$ extends to a bounded linear functional on $M_*$.

By (\ref{happy}), we have that $u$ is a bounded linear map from $A$
into $M$. As $\hat \omega$ is positive and linear, we have that $u$ is also
positive and linear. The map $u$ is unital as
\[
\langle u(1), \phi_y \rangle = \langle y, \hat \omega(1) \rangle = \langle
y, \phi \rangle = \langle 1, \phi_y \rangle.
\]
Thus $u$ carries measurement systems into measurement systems.

We now show that $vu=\hat \omega$. Let $x\in A$ and $y \in M$. We then have by
(\ref{definitionv}) and (\ref{definitionu}) that
\begin{align*}
\langle vu(x),y\rangle = \langle \phi_{u(x)},y \rangle= \langle
u(x),\phi_y \rangle = \langle \hat \omega(x),y\rangle.
\end{align*}

Let $\hat v$ denote the norm one positive linear functional on $M\cmax M$
corresponding to the map $v$. Let $\iota_{B^{**}}:B \longrightarrow
B^{**}$ denote the canonical inclusion into the double dual and
$\iota:B \longrightarrow  M$ be the map $b  \mapsto
p\iota_{B^{**}}p$.

The previous arguments show that
$$\omega(a,b)= \hat v (u(a)\otimes \iota(b)),$$
 for every $(a,b)\in A
\times B$.

\smallskip

Finally, since $\hat v$ is max continuous, there exists
a Hilbert space $H$, a state $\rho\in \mathcal S (B(H))$ and
commuting representations $\pi_1$ and $\pi_2$ of $M$ on $B(H)$ such
that,  for every  $a \in A$ and $b \in B$,
$$\omega(a,b)= \hat v (u(a)\otimes \iota(b))=
\rho(\pi_1(u(a))\pi_2(\iota(b))).$$

\section{Projections}
\label{projec}

We have shown that it is possible to simulate non-signalling locally quantum preparations of the joint system by mapping POVM elements to POVM elements via $\nu_A,\nu_B$. A natural question is then whether $\nu_A,\nu_B$ can be chosen so that, for any pair of projectors $\Pi\in A,\Pi'\in B$, $\nu_A(\Pi)$ and $\nu_B(\Pi')$ are also projective measurements. We next show that, even in finite dimensional scenarios, this is not the case.

First we state a definition which we will soon require. Given two C$^*$-algebras A and B, we say that a map $T:A \to B$ is \emph{co-completely positive} if the maps $t_n \otimes T: M_n(A) \to M_n(B)$ are positive for all $n \in \mathbb N$, where $t_n$ is the transposition map on $M_n$. It is easy to check that if $T$ is a $*$-antihomomorphism ($T(ab)=T(b)T(a)$, for all $a,b \in A$), then $T$ is co-completely positive. Similarly, if $T$ is a $*$-homomorphism, then $T$ is completely positive. Using the matrix order  defined in Subsection \ref{cpsubsection}, we can similarly define complete positivity and co-complete positivity for maps whose domains and/or ranges are given by the duals of C$^*$-algebras. It follows immediately that if $T$ is completely positive (co-completely positive, respectively), then so its adjoint $T^*$.

We assume that Alice and Bob's systems are associated to von Neumann algebras $A$ and $B$, neither of which contains $M_2(\mb C)$ as a direct summand. We assume that $\omega:P(A) \times P(B) \to [0,1]$ assigns non-signalling  probabilities to every local projective measurement system chosen by Alice and Bob. Similarly to before, using the non-signalling condition and a stronger version of Gleason's Theorem \cite[Theorem B]{BuWr}, this extends to a bounded linear map $\omega:A \times B \to \mb C$ which is positive on pure tensors. Seeking a contradiction, we also assume that $\omega$ can be simulated quantum mechanically in the commuting paradigm by mapping projections to projections; we assume that there exist C$^*$-algebras $A'$ and $B'$, a state $\rho \in (A' \otimes_{max} B')^*$, and assignments $\nu_A:P(A) \to P(A')$ and $\nu_B:P(B)\to P(B')$ that map projective measurement systems to projective measurement systems, such that $\omega(x \otimes y)=\rho(\nu_A(x) \nu_B(y))$ for all $x\in P(A)$, $y\in P(B)$. By applying the vector-valued version of Gleason's Theorem (\cite[Theorem A]{BuWr}) to $\nu_A$ and $\nu_B$, we obtain linear maps $\nu_A:A \to A'$ and $\nu_B:B \to B'$. 

In fact, the map $\nu_A$ must be a Jordan morphism, i.e., $\nu_A(x^*)=\nu(x)^*$ and $\nu_A(xy+yx)=\nu_A(x)\nu_A(y)+\nu_A(y)\nu_A(x)$. (This follows easily from the spectral theorem by approximating self-adjoint elements by finite linear combinations of mutually orthogonal projections.) We can then apply \cite[Theorem 3.3]{stormer} to get that $\nu_A$ is the sum of a $*$-homomorphism and a $*$-antihomomorphism. More precisely, there exist two orthogonal central projections $E_A^1$ and $E_A^2$ in 
$\overline{\nu_A(A)}^{weak-*}$ such that  
$\nu_{A}^1: x \mapsto \nu_A(x)E_A^1$  is a $*$-homomorphism, and  
$\nu_{A}^2: x \mapsto \nu_{A}(x)E_A^2$ is a $*$-antihomomorphism, with 
$E_A^1 + E_A^2 = 1$ and $\nu_A=\nu_A^1 +\nu_A^2$.
Similarly, there exist two orthogonal central projections $E_B^1$ and $E_B^2$ in $\overline{\nu_B(B)}^{weak-*}$ such that the map $\nu_{B}^1: x \mapsto \nu_B(x)E_B^1$ is a $*$-homomorphism and that the map $\nu_{B}^2: x \mapsto \nu_{B}(x)E_B^2$ is a $*$-antihomomorphism, $E_B^1 + E_B^2 = 1$, and $\nu_B=\nu_B^1 +\nu_B^2$ as linear maps.

We recall that we can associate $\omega:A \times B \to \mathbb C$ with a positive linear map $\hat \omega: A \to B^*$. Similarly, we can associate $\rho \in (A' \otimes_{max} B')^*$ with a positive linear map $\hat \rho:A' \to (B')^*$. Indeed, by Theorem \ref{PisierTheorem}, the map $\hat \rho$ is completely positive. We then have the following decomposition of $\hat \omega$:
\[
\xymatrix{
A \ar[r]^{\hat \omega} \ar[d]_{\nu_A^1 + \nu_A^2} & B^* \\
A' \ar[r]^{\hat \rho} & (B')^* \ar[u]_{(\nu_B^1)^*+(\nu_B^2)^*}\\
}.
\]
That is,
\[
\hat \omega = (\nu_B^1)^* \hat \rho \nu_A^1 + (\nu_B^2)^*\hat \rho \nu_A^1 
+ (\nu_B^1)^*\hat \rho \nu_A^2 + (\nu_B^2)^* \hat \rho \nu_A^2.
\]
We have now written $\hat \omega$ as the sum of completely positive and co-completely positive maps. As the composition of completely positive maps, it is clear that $(\nu_B^1)^* \hat \rho ^*\nu_A^1$ is completely positive. As the maps $\nu_A^2$ and $(\nu_B^2)^*$ are co-completely positive, the maps $t_n \otimes \nu_A^2$ and $t_n \otimes (\nu_B^2)^*$ are positive for each $n$, and thus
\[
id_n \otimes (\nu_B^2)^* \hat \rho \nu_A^2 =(t_n \otimes (\nu_B^2)^*)(id_n \otimes \hat \rho)(t_n \otimes \nu_A^2)
\]
is positive for each $n \in \mathbb N$. Thus the map $(\nu_B^2)^* \hat \rho \nu_A^2$ is completely positive. Similarly, the maps $(\nu_B^2)^*\hat \rho \nu_A^1$ and $(\nu_B^1)^*\hat \rho \nu_A^2$ are co-completely positive.

However, not every positive linear map is decomposable in this fashion. 
 Appendix B in \cite{Choi} provides an explicit counterexample in the form of a map $\phi:M_3(\mb C) \to M_3(\mb C)$ that cannot be written as the sum of completely positive and co-completely positive maps. This contradiction shows that it is not, in general, possible to simulate locally quantum non-signalling distributions while mapping projective measurement systems to projective measurement systems.
 
\section*{Acknowledgements}
M.N. acknowledges support by the Templeton Foundation and the European Commission (Integrated Project QESSENCE). This work was supported by the Spanish grants I-MATH, MTM2011-26912, QUITEMAD and the European project QUEVADIS.

\end{document}